\documentclass[a4paper,11pt]{article}
\pdfoutput=1 % if your are submitting a pdflatex (i.e. if you have
\usepackage{jheppub} % for details on the use of the package, please
\usepackage[T1]{fontenc} % if needed
\usepackage[italian,english]{babel}
\usepackage{hyperref}
\usepackage{ifpdf}
\usepackage{subfigure}
\usepackage{tikz}
\usepackage[compat=1.1.0]{tikz-feynman}
\usepackage{slashed}
\usetikzlibrary{arrows,shapes}
\usetikzlibrary{trees}
\usetikzlibrary{matrix,arrows} % For commutative diagram
\usetikzlibrary{positioning}% For "above of=" commands
\usetikzlibrary{calc,through}% For coordinates
\usetikzlibrary{decorations.pathreplacing}  % For curly braces
\usepackage{pgffor}% For repeating patterns

\usetikzlibrary{decorations.pathmorphing}% For Feynman Diagrams
\usetikzlibrary{decorations.markings}
\tikzset{
vector/.style={decorate, decoration={snake}, draw},
fermion/.style={draw=black, postaction={decorate}}, 
scalar/.style={dashed,draw=black, postaction={decorate}}}
\tikzstyle{block} = [draw, rectangle, 
minimum height=3em, minimum width=6em]

\usepackage{amssymb}
\usepackage{amsfonts}
\usepackage{epsf}
\usepackage{rotating}
\usepackage{graphicx}
\usepackage{amsmath}
\usepackage{fancyhdr}
\usepackage{lineno}
\usepackage{babel}
\usepackage{graphics}
\usepackage{pstricks}
\usepackage{color}
\usepackage{multirow}
\usepackage{float}
\usepackage{lineno}
\usepackage{mathtools}
\usepackage{stackengine}
\usepackage{comment}
\newcommand{\lsim}{\mathrel{\mathop{\kern 0pt \rlap
{\raise.2ex\hbox{$<$}}}
\lower.9ex\hbox{\kern-.190em $\sim$}}}
\newcommand{\gsim}{\mathrel{\mathop{\kern 0pt \rlap
{\raise.2ex\hbox{$>$}}}
\lower.9ex\hbox{\kern-.190em $\sim$}}}

\newcommand{\be}{\begin{equation}}
\newcommand{\ee}{\end{equation}}
\newcommand{\bea}{\begin{eqnarray}}
\newcommand{\eea}{\end{eqnarray}}

\def\gev{\ensuremath{\mathrm{\,Ge\kern -0.1em V\,}}}
\def\tev{\ensuremath{\mathrm{\,Te\kern -0.1em V\,}}}

\begin{document}

\title{Pseudo-scalar dark matter from a broken gauged symmetry }

\author[a]{Junho Kang}
\author[b]{, Sarif Khan}
\author[c]{, Jongkuk Kim}
\author[a]{, and Hyun Min Lee}

\affiliation[a]{Department of Physics, Chung-Ang University, Seoul 06974, Korea.}
\affiliation[b]{Department of Physics, Aliah University, Kolkata - 700014, India.}
\affiliation[c]{Division of Liberal Studies, Kangwon National University, Samcheok 25913,  Korea.}

\emailAdd{junobonnie12@gmail.com}
\emailAdd{skhan.phys@aliah.ac.in}
\emailAdd{jongkuk.kim927@gmail.com}
\emailAdd{hminlee@cau.ac.kr}

\keywords{} 

\abstract{We propose a novel model for pseudo-scalar dark matter (PSDM) by extending the Standard Model (SM) with a dark gauged $U(1)_X$ symmetry, but without dark charged fermions. 
We impose a $Z_2$ symmetry to ensure the stability of pseudo-scalar dark matter and regard the  $U(1)_X$ symmetry as being broken dominantly by a large VEV of the singlet scalar field. The would-be Goldstone associated with the $U(1)_X$ gauge boson is almost orthogonal to the direction of PSDM. 
As a result, we show that PSDM appears as a stable pseudo-Nambu-Goldstone boson receiving the mass from the $U(1)_X$ invariant mixing potential and the corresponding cross section for direct detection gets suppressed even for the weak-scale mass of PSDM. We also show that the correct relic density can be explained by the PSDM annihilations into the SM particles  or into a pair of light Higgs-like scalars, being compatible with the bounds from Higgs invisible decay, Higgs data and indirect detection.   

}

\maketitle              

\section{Introduction}

Weakly Interacting Massive Particles (WIMPs) are good candidates for dark matter, being motivated by various solutions to the hierarchy problem for the Higgs mass in the Standard Model, so providing an alternative testing ground for new physics at the weak scale in addition to collider searches for colored partners of the SM such as Large Hadron Collider (LHC).
However, the minimal models for WIMP have been strongly constrained due to null results in direct detection experiments and there is no convincing evidence for new colored particles at the LHC either.

Suppose that WIMP is a pseudo-Nambu-Goldstone boson(pNGB) or a pseudo-scalar as remnants of new global symmetries in the dark sector.
It could appear as dark pions from the strong dynamics of dark QCD or it could be the consequence of a spontaneously broken global $U(1)$ symmetry in the dark sector. The common feature of pNGB dark matter is that it has derivative interactions to the SM due to shift symmetries, which are broken only by mass terms. Examples include QCD axion or axion-like dark matter, which could couple to photons or gluons by chiral anomalies, so they would be a decaying dark matter with a sufficiently long lifetime longer than the age of the universe \cite{axionDM}. In the case of dark pions, dark flavor symmetries, if broken by strong dynamics without dark chiral anomalies, could protect dark matter from decaying \cite{darkpions}. 

In this article, we propose a new model with dark gauged $U(1)_X$ symmetry for pseudo-scalar dark matter where there are no light dark fermions charged under the $U(1)_X$. We identify the $U(1)_X$-neutral combination of pseudo-scalars as pseudo-scalar dark matter. We ensure the stability of dark matter by a $Z_2$ symmetry surviving even after the dark $U(1)_X$ symmetry is broken spontaneously at a high scale. We obtain the mass and the non-derivative couplings for dark matter by a renormalizable or non-renormalizable dark sector potential respecting the dark $U(1)_X$ symmetry but violating an approximate global $U(1)$ symmetry associated with pseudo-scalar dark matter. We remark that the axion quality can be guaranteed by similar mechanisms such as discrete $R$ symmetries, which are originated from the isometry of extra dimensions,  because of the protection of the global $U(1)$ Peccei-Quinn symmetry at sufficiently higher orders \cite{Rsymm}.

We find that  there is a cancellation between derivative and non-derivative couplings in the DM direct detection cross section in the polar basis for scalar fields \cite{PSDM}, so weak-scale pseudo-scalar dark matter is compatible with both direct detection bounds and the correct relic density. We also check the robustness of the cancellation mechanism for generic dark Higgs couplings for pseudo-scalar dark matter. 
There was a similar model for pseudo-scalar dark matter with a $U(1)_{B-L}$ symmetry \cite{BL}, but it was recognized that pseudo-scalar dark matter is unstable in this case, due to the fact that the SM fermions and right-handed neutrinos carry nonzero $U(1)_{B-L}$ charges.

The paper is organized as follows.
We begin with a model setup including the matter content and the $U(1)_X$ invariant Lagrangian.
Then, we compute the mass spectrum and mixings for scalar fields after the $U(1)_X$ symmetry is broken by the VEVs of the singlet scalar fields. We also identify both derivative and non-derivative interactions for pseudo-scalar fields in the polar basis. Next we consider the constraints from direct detection, relic density, indirect detection and Higgs invisible decay. 
Finally, conclusions are drawn.
There are three appendices connecting the physical masses and mixings to the quartic couplings in the scalar potential, the trilinear couplings between pseudo-scalar DM and the CP-even scalars in the model, and the annihilation cross sections for pseudo-scalar DM, respectively.

\section{The setup}
\label{setup}

We set up the model for pseudo-scalar dark matter in the extension of the SM with an extra gauged $U(1)_X$ symmetry. 
We introduce two complex scalar fields, $\Phi$ and $S$, which are neutral under the SM gauge groups but carry nonzero $U(1)_X$ charges, and the SM Higgs $H$ and the SM fermions, as summarized in Table~\ref{charges}. We also have an extra gauge boson $X_\mu$, associated with the $U(1)_X$.

\begin{table}[hbt!]
\begin{center}
\scalebox{0.9}{
    \begin{tabular}{||c|c|c|c|c|c|c|c|c||}
      \hline\hline 
      & $Q^i_L={\scriptsize\left(\begin{array}{c}u^i_L\\ d^i_L \end{array}\right)}$ & $u^i_{R}$  &  $d^i_{R}$ & $L^i_{L}={\scriptsize\left(\begin{array}{c}\nu^i_L\\ e^i_L \end{array}\right)}$  & $e^i_{R}$
      & $H$ &  $\Phi$ & $S$   \\
      \hline
      $G_{\rm SM}$ & $(3,2)_{+\frac{1}{6}}$ & $(3,1)_{+\frac{2}{3}}$ & $(3,1)_{-\frac{1}{3}}$
            & $(1,2)_{-\frac{1}{2}}$ & $(1,1)_{-1}$ & $(1,2)_{+\frac{1}{2}}$ & $(1,1)_0$ & $(1,1)_0$  \\
                      \hline
            $U(1)_X$ & $0$ & $0$ & $0$
            & $0$ & $0$ & $0$  & $-\frac{2}{n}$ & $1$  \\ 
      \hline\hline
    \end{tabular}} 
      \end{center}
    \caption{Representations under the SM and $U(1)_X$ gauge groups.
    \label{charges}}
\end{table}

The Lagrangian of the model is given by
\begin{eqnarray}
\mathcal{L}& = &\mathcal{L}_{SM} + 
\left(D^{\mu} \Phi \right)^{\dagger} \left( D_{\mu} \Phi \right) +\left(D^{\mu} S \right)^{\dagger} \left( D_{\mu} S \right) 
+ \mathcal{L}_{U(1)_X}
 - V\left(H,\Phi, S \right)\,,
\end{eqnarray}
with
\begin{eqnarray}
V(H,\Phi, S) & = &
- \mu^{2}_H \left( H^\dagger H \right) +
\lambda_H \left( H^\dagger H\right)^{2}
- \mu^{2}_\Phi \left( \Phi^\dagger\Phi \right) +
\lambda_\Phi \left(\Phi^\dagger\Phi  \right)^{2}
- \mu^{2}_S S^\dagger S 
\nonumber \\
&&+
\lambda_S \left( S^\dagger S\right)^{2} + \lambda_{H\Phi} \left(H^\dagger H \right)
\left( \Phi^\dagger \Phi \right)
+ \lambda_{HS} \left( H^\dagger H \right)
\left( S^\dagger S \right) \nonumber \\
&&+ \lambda_{\Phi S} \left(\Phi^\dagger \Phi\right)
\left( S^\dagger S \right)
-\left( \frac{\kappa_n}{\Lambda^{n-2}}  \Phi^n S^2 
+  {\rm h.c.} \right). \label{pot}
\end{eqnarray}
Here, the covariant derivatives for the singlet scalar fields are given by
\bea
D_\mu S= \Big( \partial_\mu-i g_X X_\mu\Big)S, \quad D_\mu \Phi= \bigg( \partial_\mu+i \frac{2}{n} g_X X_\mu\bigg)\Phi,
\eea
and $\Lambda$ is of order the largest value of the mass parameters in the theory for $n=1$ and the cutoff scale for $n\geq 3$.
We note that the above Lagrangian has the $Z_2$ invariance under 
\bea
S\to S^*, \quad \Phi\to \Phi^*, \quad  X_\mu \to - X_\mu, \label{Z2}
\eea
as far as $\kappa_n$ is a real parameter \footnote{Even if $\kappa_n$ is a complex parameter in the original basis for fields, we can always rotate away a nonzero CP phase of $\kappa_n$ by the field redefinitions of $\Phi$ and/or $S$. This corresponds to the trivial tadpole condition in the vacuum \cite{ligong}.}, 
so the lightest of ${\rm Im} \,S,  {\rm Im}\, \Phi$ and $X$ can be a dark matter candidate \footnote{A similar model with $n=1$ in the case of the $B-L$ gauge symmetry was considered \cite{BL}, but pseudo-scalar would be unstable because the $Z_2$ symmetry is broken explicitly by the $B-L$ gauge interactions to the SM fermions.}. 
In particular, the linear combination of CP-odd scalars can be a pseudo-scalar dark matter, if it is lighter than the $X$ gauge boson.
Moreover, the above Lagrangian is invariant under an additional $Z'_2$ symmetry under which
\bea
S\to -S, \quad \Phi\to \Phi, \quad X_\mu\to X_\mu.
\eea
Thus, the $Z'_2$-odd terms such as $\Phi S$ for $n=2$ and $\Phi^2 S$ for $n=4$ are forbidden. On the other hand, the $Z'_2$ symmetry is not necessary for $n=1$ and $n=3$ cases. If it were not for the $Z'_2$ symmetry, the $Z'_2$-odd terms such as $\Phi S$ for $n=2$ and $\Phi^2 S$ for $n=4$ are allowed, so the direct detection cross section could not be protected due to the extra contributions from the effective linear terms in $S$ after $\Phi$ gets a VEV.

We remark that higher terms containing the $U(1)_X$ invariant combination, $\Phi^n S^2$, such as $\frac{1}{\Lambda^{n}}|\Phi|^2(\Phi^n S^2)$,  $\frac{1}{\Lambda^{2n}}(\Phi^n S^2)^2$, etc, and their complex conjugates, can be also introduced, but they are suppressed by extra powers of the cutoff scale, so the following discussion on the direct detection of pseudo-scalar dark matter would be affected little. However, we note that the coefficients of such $U(1)_X$-invariant higher order terms would give rise to extra CP phases in general. Thus, the extra CP phases must be suppressed up to the sufficiently higher orders for the $Z_2$ symmetry to remain as a good symmetry for the protection of PSDM.

%%%%%%%%%%%%%%%%%%%%%%%%%%%%%%%%%%%%%%%%%%%%%%%%%%%%%%%%%%%%%%%
\section{Mass spectrum and pseudo-scalar interactions}
\label{masses}
%%%%%%%%%%%%%%%%%%%%%%%%%%%%%%%%%%%%%%%%%%%%%%%%%%%%%%%%%%%%%%%

We consider the mass spectrum and mixings for scalar fields in the model and identify the interactions for pseudo-scalar dark matter in the model.

\subsection{Scalar mass spectrum and mixings}

The electroweak and $U(1)_X$ gauge symmetries are broken spontaneously when the scalar fields acquire vacuum expectation values as follows,
\begin{eqnarray}
H =
\begin{pmatrix}
G^{+} \\
\frac{v+h + i G^{0}}{\sqrt{2}} 
\end{pmatrix},\quad
S = \frac{v_S + \rho_S + i A_S  }{\sqrt{2}},\quad \Phi= \frac{v_\Phi + \rho_\Phi + i A_\Phi  }{\sqrt{2}}.
\label{scalars}
\end{eqnarray} 
After the symmetry breaking, the $Z_2$ symmetry in eq.~(\ref{Z2}) remains unbroken, appearing as $A_S\to -A_S$,  $A_\Phi\to -A_\Phi$ and $X_\mu\to -X_\mu$. We also note that the $U(1)_X$ symmetry is broken to a discrete $Z_2$ symmetry for $n\geq 2$,  due to the VEVs of the singlet scalar fields, as can be seen by rescaling the $U(1)_X$ charges of the singlet scalar fields by $n$ in Table~\ref{charges}.  

The model contains three physical CP-even Higgs states and one physical CP-odd scalar, while the remaining scalar degrees of freedom provide the longitudinal modes of the massive gauge bosons for SM and $U(1)_X$ gauge symmetries.  
The minimization of the potential {\it i.e.} 
$\left( \frac{\partial \mathcal{V}}{\partial \phi_i} \right)_{\psi_i=0} = 0$ where $\phi_i = h, \rho_\Phi, \rho_S$, and $\psi_i$ are the remaining scalar fields, give rise to
\begin{eqnarray}
&& \mu^2_H= \frac{1}{2} \left( 2 \lambda_H v^2 + \lambda_{H\Phi} v^2_\Phi + \lambda_{H S} v^2_S \right), \nonumber \\
&& \mu^2_S = \frac{1}{2} \left( \lambda_{HS} v^2 + 2 \lambda_S v^2_S 
+ \lambda_{\Phi S} v^2_\Phi - 2^{2-\frac{n}{2}}  \frac{\kappa_n}{\Lambda^{n-2}}  v^n_\Phi  \right), \nonumber \\
&& \mu^2_\Phi = \frac{1}{2} \left( \lambda_{H\Phi} v^2 + 2 \lambda_\Phi v^2_\Phi 
+ \lambda_{\Phi S} v^2_S - 2^{1-\frac{n}{2}} \frac{\kappa_n}{\Lambda^{n-2}} n v^{n-2}_\Phi  v^2_S\right).
\end{eqnarray}
Then, using the above tadpole conditions, we obtain the mass matrix for CP-even scalars in the basis $(h,\rho_S, \rho_\Phi)$, as follows,
\begin{eqnarray}
{\cal M}^2_H = 
\begin{pmatrix}
2 \lambda_H v^2 & \lambda_{H S} v v_S & \lambda_{H\Phi} v v_\Phi \\
\lambda_{H S} v v_S  & 2 \lambda_S v^2_S 
& v_S \left(  \lambda_{\Phi S} v_\Phi - 2^{1-\frac{n}{2}} \frac{\kappa_n}{\Lambda^{n-2}} n v_\Phi^{n-1}\right) \\
\lambda_{H\Phi} v v_\Phi &  v_S \left(  \lambda_{\Phi S} v_\Phi - 2^{1-\frac{n}{2}} \frac{\kappa_n}{\Lambda^{n-2}} n v_\Phi^{n-1}\right) & 2 \lambda_\Phi v^2_\Phi  -2^{-\frac{n}{2}}   \frac{\kappa_n}{\Lambda^{n-2}}  (n-2) n v^2_S v^{n-2}_\Phi
\end{pmatrix}.
\label{CPeven}
\end{eqnarray}

The mass eigenstates $(h_{1},\,\,h_{2},\,\,h_{3})$ and the interaction eigenstates 
$(h\,,\,\rho_S, \,\,\rho_\Phi)$ of the CP-even scalar sector are related by
\begin{eqnarray}
\begin{pmatrix}
h\\
\rho_S\\
\rho_\Phi
\end{pmatrix}
= R
\begin{pmatrix}
h_{1}\\
h_{2}\\
h_{3}
\end{pmatrix}
\label{U-PMNS}
\end{eqnarray} 
where $R$ is the $3\times 3$ orthogonal rotation matrix, given by
\bea
R=\begin{pmatrix}
c_{12} c_{13} & s_{12} c_{13} & s_{13} \\
-s_{12} c_{23} - c_{12} s_{23} s_{13} & c_{12} c_{23} - s_{12} s_{23} s_{13}
& s_{23} c_{13} \\
s_{12} s_{23} - c_{12} c_{23} s_{13} & -c_{12} s_{23} - s_{12} c_{23} s_{13}
& c_{23} c_{13}   
 \end{pmatrix}.
 \label{mixing}
\eea
Here, $c_{12}\equiv \cos\theta_{12}, s_{12}=\sin\theta_{12}$, etc, and $\theta_{12}, \theta_{23}, \theta_{13}$ are the mixing angles between $h$ and $\rho_S$, between $\rho_S$ and $\rho_\Phi$ and between $h$ and $\rho_\Phi$, respectively.

Similarly, the mass matrix for CP-odd scalars  in the basis $(A_\Phi , A_S)$ is  given by
\begin{eqnarray}
{\cal M}^2_A =\frac{\kappa_n}{\Lambda^{n-2}}
\begin{pmatrix}
  2^{-\frac{n}{2}} n^{2} v_S^2 v^{n-2}_\Phi & 2^{1-\frac{n}{2}}  n v_S v^{n-1}_\Phi \\
 2^{1-\frac{n}{2}}  n v_S v^{n-1}_\Phi &   2^{2-\frac{n}{2}}  v^n_\Phi 
\end{pmatrix}.
\end{eqnarray}
Then, the mass eigenvalues for CP-odd scalars are given by
\begin{eqnarray}
M_{G} = 0,\quad M^2_{A} = 2^{2-\frac{n}{2}}\, \frac{\kappa_nv^n_\Phi }{\Lambda^{n-2}} 
\left( 1 + \frac{n^{2} v^2_S}{4 v^2_\Phi} \right)\,. \label{Amass}
\end{eqnarray}
and the mass eigenstates and interaction eigenstates of the CP-odd scalars are related by
\begin{eqnarray}
\begin{pmatrix}
G \\
A
\end{pmatrix}
=
\begin{pmatrix}
 \cos\zeta & -\sin\zeta \\
 \sin\zeta & \cos\zeta
\end{pmatrix}
\begin{pmatrix}
A_\Phi \\
A_S
\end{pmatrix}
\end{eqnarray}
where $\tan\zeta = \frac{n v_S}{2 v_\Phi}$.
Here, $G$ is the would-be Goldstone boson appearing after the $U(1)_X$ is broken spontaneously, while the mass of the $U(1)_X$ gauge boson $X$ is given by
\begin{eqnarray}
M^2_X &=&  \frac{4 g^2_X v^2_\Phi }{n^2} + g^2_X v^2_S,\nonumber \\
 &=& \frac{4g^2_Xv^2_\Phi}{n^2} \left(1 + \tan^{2}\zeta \right).
\end{eqnarray}
Using the results, we can rewrite the mass matrix for CP-even scalars in eq.(\ref{CPeven}) as
\begin{eqnarray}
{\cal M}^2_H = 
\begin{pmatrix}
2 \lambda_H v^2 & \lambda_{H S} v v_S & \lambda_{H\Phi} v v_\Phi \\
\lambda_{H S} v v_S  & 2 \lambda_S v^2_S 
&   \lambda_{\Phi S} v_\Phi v_S - \frac{n v_S}{2 v_\Phi \left( 1 + \frac{n^{2} v^2_S}{4 v^2_\Phi} \right)}  M^2_A \\
\lambda_{H\Phi} v v_\Phi &  \lambda_{\Phi S} v_\Phi v_S - \frac{n v_S}{2 v_\Phi \left( 1 + \frac{n^{2} v^2_S}{4 v^2_\Phi} \right)} M^2_A & 2 \lambda_\Phi v^2_\Phi  - \frac{n(n-2) v^2_S}{4v_\Phi^2 \left( 1 + \frac{n^{2} v^2_S}{4 v^2_\Phi} \right)} M^2_A
\end{pmatrix}.
\label{CPeven2}
\end{eqnarray}

In the limit of decoupling the $\Phi$ dark Higgs sector with $v_\Phi \gg v_S, v$, the masses for the $X$ gauge boson and the pseudo-scalar mass  are approximated to
\bea
M^2_X&\simeq & \frac{4g^2_X v^2_\Phi}{n^2}, \\
M^2_A &\simeq&   2^{2-\frac{n}{2}}\, \frac{\kappa_nv^n_\Phi }{\Lambda^{n-2}}.
\eea
In this case, we can make the pseudo-scalar dark matter light  for appropriate choices of $v_\Phi$, $\kappa_n$ and $\Lambda$.  
For instance, we get $M^2_A\simeq 2\sqrt{2} \kappa_1 v_\Phi \Lambda$ for $n=1$; $M^2_A\simeq 2\kappa_2 v^2_\Phi$ for $n=2$; $M^2_A\simeq \sqrt{2} \kappa_3 \frac{v^3_\Phi}{\Lambda}$ for $n=3$;  $M^2_A\simeq  \kappa_4 \frac{v^4_\Phi}{\Lambda^2}$ for $n=4$, etc. For the pseudo-scalar mass below the weak scale for $v_\Phi\gg v_S, v$, we would need $\kappa_1\lesssim \frac{v^2}{v_\Phi \Lambda}$ for $n=1$; $\kappa_2\lesssim \frac{v^2}{v^2_\Phi}$ for $n=2$; $\kappa_3\lesssim \frac{v^2\Lambda}{v^3_\Phi}$ for $n=3$; $\kappa_4\lesssim \frac{v^2\Lambda^2}{v^4_\Phi}$ for $n=4$, etc. Then, for both $n=1$ and $n=2$, we would need a very small $\kappa_1$ or $ \kappa_2$. But, for $n=3$ or $n=4$, we can take $\kappa_3$ and $ \kappa_4$ to be of order one as far as $\Lambda\sim v_\Phi (v_\Phi/v)^2$ or $\Lambda\sim v_\Phi (v_\Phi/v)$.

From eq.~(\ref{CPeven2}), the CP-even scalar $\rho_\Phi$ gets decoupled from the Higgs and the CP-even scalar $\rho_S$, for $v_\Phi\gg v_S, v$ and $\lambda_{\Phi S}, \lambda_{H\Phi}\lesssim \lambda_\Phi$. Therefore, for $v_\Phi\gg v_S, v$, we can obtain the effective theory only for $h, A, \rho_S$, while $X, \rho_\Phi$ fields are decoupled. 

We note that we can still accommodate a sizable mixing between the SM Higgs and $\rho_S$ for $v_S\sim v$, as in the original model for pseudo-scalar dark matter \cite{PSDM}. However, in our case, the form of the pseudo-scalar interactions are protected by the $U(1)_X$ symmetry against the loop corrections unlike the case without a protecting symmetry in Ref.~\cite{PSDM}.

\subsection{Pseudo-scalar interactions in the polar basis}

For direct detection, it is more convenient to work in the polar basis for singlet scalar fields, so we choose the singlet scalar fields in our model, as follows,
\bea
S=\frac{1}{\sqrt{2}} (v_S+\rho_S) \,e^{ia_S/v_S}, \quad \Phi=\frac{1}{\sqrt{2}} (v_\Phi+\rho_\Phi) \,e^{ia_\Phi/v_\Phi},
\eea
instead of the linear representations in eq.~(\ref{scalars}). Then, the scalar potential in eq.~(\ref{pot}) becomes
\bea
V(h,\rho_\Phi, \rho_S, a_S, a_\Phi)&=&-\frac{1}{2} \mu^2_H {\bar h}^2 +\frac{1}{4}\lambda_H {\bar h}^4 -\frac{1}{2}\mu^2_\Phi {\bar \rho}^2_\Phi +\frac{1}{4}\lambda_\Phi {\bar\rho}^4_\Phi -\frac{1}{2}\mu^2_S {\bar\rho}^2_S \nonumber \\
&&+ \frac{1}{4}\lambda_S {\bar\rho}^4_S +\frac{1}{4}\lambda_{H\Phi}{\bar h}^2 {\bar \rho}^2_\Phi +\frac{1}{4} \lambda_{HS} {\bar h}^2 {\bar\rho}^2_S \nonumber \\
&&+\frac{1}{4} \lambda_{\Phi S} {\bar\rho}^2_\Phi {\bar\rho}^2_S - 2^{-\frac{n}{2}}\frac{\kappa_n}{\Lambda^{n-2}}\,{\bar\rho}^n_\Phi {\bar\rho}_S^2\cos\bigg(\frac{n a_\Phi}{v_\Phi}+ \frac{2a_S}{v_S}\bigg)
\eea
where ${\bar h}=h+v, {\bar\rho}_\Phi=v_\Phi + \rho_\Phi$ and ${\bar\rho}_S=v_S+\rho_S$.
A combination of pseudo-scalars, $A\sim \frac{n a_\Phi}{v_\Phi}+ \frac{2a_S}{v_S}$, becomes massive, while the orthogonal combination, namely,  $G\sim -\frac{2 v_\Phi}{n}  a_\Phi +  v_S a_S$, corresponds to the would-be Goldstone boson associated with the broken $U(1)_X$ gauge symmetry, in agreement with our discussion in the linear basis in the previous section.
As a result, the interactions for CP-odd scalars in the potential appear only through $\kappa_n$.

On the other hand, the kinetic terms for singlet scalar fields lead to 
\bea
{\cal L}_{\rm kin}&\supset &\frac{1}{2} (v_\Phi +\rho_\Phi)^2 \bigg(\frac{\partial_\mu a_\Phi }{v_\Phi}+\frac{2}{n} g_X X_\mu\bigg)^2 + \frac{1}{2} (v_S +\rho_S)^2 \bigg(\frac{\partial_\mu a_S}{v_S}-g_X X_\mu\bigg)^2 \nonumber \\
&&=\frac{1}{2} \bigg(1 +\frac{\rho_\Phi}{v_\Phi}\bigg)^2(\partial_\mu a_\Phi )^2+\frac{1}{2} \bigg(1 +\frac{\rho_S}{v_S}\bigg)^2(\partial_\mu a_S )^2 \nonumber \\
&&+g_X\bigg((v_\Phi +\rho_\Phi)^2\frac{2}{n } \frac{\partial^\mu a_\Phi}{v_\Phi}-  (v_S +\rho_S)^2\frac{\partial^\mu a_S}{v_S} \bigg) X_\mu \nonumber \\
&&+\frac{1}{2}g^2_X  \bigg(\frac{4}{n^2}(v_\Phi +\rho_\Phi)^2+  (v_S +\rho_S)^2 \bigg) X_\mu X^\mu.
\eea 

We now choose the unitary gauge for $U(1)_X$, namely, $G=0$ or $a_\Phi=\frac{n v_S}{2v_\Phi}\, a_S$.
As a result, we find the part of the Lagrangian containing the remaining pseudo-scalar $a_S$, as follows,
\bea
{\cal L}_{a_S}={\cal L}_{a_S,{\rm kin}}+{\cal L}_{a_S, 1}+{\cal L}_{a_S, 2}
\eea
with
\bea
{\cal L}_{a_S,{\rm kin}}&=&\frac{1}{2} \bigg(1+\frac{n^2 v^2_S}{4 v^2_\Phi}\bigg) (\partial_\mu a_S)^2+\frac{1}{2} \bigg(\frac{2\rho_S}{v_S}+ \frac{\rho^2_S}{v^2_S} + \frac{n^2 v^2_S \rho_\Phi}{2 v^3_\Phi}+\frac{n^2 v^2_S \rho^2_\Phi}{4 v^4_\Phi}\bigg) (\partial_\mu a_S)^2, \label{int1} \\
{\cal L}_{a_S, 1} &=& -2^{-\frac{n}{2}}\frac{\kappa_n}{\Lambda^{n-2}}\,(v_\Phi+\rho_\Phi)^n(v_S+\rho_S)^2\cos\bigg(\bigg(1+\frac{n^2 v^2_S}{4 v^2_\Phi}\bigg) \frac{2a_S}{v_S}\bigg), \label{int2} \\
{\cal L}_{a_S, 2} &=&g_X v_S \bigg(\frac{2\rho_\Phi}{v_\Phi} +\Big(\frac{\rho_\Phi}{v_\Phi}\Big)^2- \frac{2\rho_S}{v_S} -\Big(\frac{\rho_S}{v_S}\Big)^2\bigg) \partial^\mu a_S X_\mu. \label{int3}
\eea
Here, we note that the resultant kinetic term for $a_S$ is not a canonical form, so we need to redefine the pseudo-scalar field by
\bea
A = \bigg(1+\frac{n^2v^2_S}{4v^2_\Phi}\bigg)^{\frac{1}{2}} a_S.
\eea
Then, we find that the mass for $A$ is the same as  in the linear basis in eq.~(\ref{Amass}), because
\bea
M^2_A= \bigg(1+\frac{n^2v^2_S}{4v^2_\Phi}\bigg)^{-1} \frac{\partial^2 V}{\partial a^2_S}\bigg|_{a_S=0},
\eea
and we can also identify the interaction terms for $A$ from eqs.~(\ref{int1})-(\ref{int3}).

We make several remarks in order.
In particular,  both eqs.~(\ref{int1}) and (\ref{int2}) contain the interaction terms for pseudo-scalar dark matter that are relevant for direct detection at tree level, such as $A-\rho_S, A^2-\rho_\Phi$, in the presence of the Higgs mixings with $\rho_S$ and/or $\rho_\Phi$. The protection mechanism for direct detection can be seen clearly in the polar basis, as the nontrivial interactions for pseudo-scalar dark matter appear only from the $\kappa_n$ term, which breaks a $U(1)$ global symmetry associated with $A$ explicitly.
In the next subsection, we will discuss the cross section for direct detection in the polar basis for pseudo-scalar dark matter in detail.

Moreover, there are interaction terms between the pseudo-scalar, the CP-even singlet scalars  and the $X$ gauge boson in eq.~(\ref{int3}). For instance, the cubic gauge interactions such as $A-X-\rho_S$ and $A-X-\rho_\Phi$ are responsible for the decays of the $X$ gauge boson and the quartic gauge interactions such as $A-\rho^2_S-X$ and $A-\rho^2_\Phi-X$, so they can lead to new annihilation channels for pseudo-scalar dark matter if $X, \rho_\Phi$ are not decoupled completely.

\subsection{Higgs-like interactions to the SM}

In the presence of the general mixings between CP-even scalars in eq.~(\ref{mixing}), the Yukawa interactions of the CP-even scalars to the SM fermions $f$ are
\bea
{\cal L}_{\rm Yukawa}=-\sum_f \sum_{i=1}^3 \frac{m_f}{v}\, R_{1i}h_i  {\bar f} f.
\eea
Similarly, the interactions between the CP-even scalars and $W, Z$ bosons are similarly modified to
\bea
{\cal L}_{\rm gauge}=\delta_V M^2_V\bigg[\bigg(1+\frac{1}{v} \sum_{i=1}^3R_{1i}h_i\bigg)^2-1\bigg] V_\mu V^\mu
\eea
where $V= W, Z$ and $\delta_V=1, \frac{1}{2}$ for $V=W, Z$.

\section{Constraints on the model}

We impose various constraints on the model coming from direct detection experiments, dark matter relic density, indirect detection as well as Higgs data and collider bounds.

\subsection{Dark matter direct detection}

The CP-even scalars mediate between the pseudo-scalar DM and the SM quarks for direct detection.  
So, we find the  spin-independent cross section for the DM-nucleon scattering, as follows,
\bea
\sigma^{SI}_{A-N}=\frac{\mu^2_N}{\pi M^2_A A^2} \bigg[Z f_p + (A-Z)f_n \bigg]^2
\eea
where  $Z, A-Z$ are the numbers of protons and neutrons in the target nucleus, respectively, $\mu_N=m_N  M_A/(m_N+M_A) $ is the reduced mass for the DM-nucleon system, and $f_{p, n}$ are the nucleon form factors, given by
\bea
f_N=\frac{m_N v_S}{v}\, {\tilde\lambda}\bigg(\sum_{q=u,d,s}\,f^N_{Tq}+ \frac{2}{9} f^N_{TG} \bigg),\quad N=p,n,
\eea
with  $f^N_{TG}=1-\sum_{q=u,d,s} f^N_{Tq}$, and the effective DM coupling being
\bea
{\tilde\lambda}=-\sum_{i=1}^3 R_{1i} (R_{2i} A_S+ R_{3i} A_\Phi)\, \frac{1}{m^2_{h_i}}. \label{DMeff}
\eea
Here,  $f^N_{Tq}$ is the mass fraction of quark $q$ inside the nucleon $N$, defined by $\langle N|m_q {\bar q}q |N\rangle= m_N f^N_{Tq}$, and $f^{N}_{TG}$ is the mass fraction of gluon $G$ inside  the nucleon $N$, due to heavy quarks \cite{hisano}. We quote the updated numerical values as $f^p_{T_u}=0.0208\pm 0.0015$ and $f^p_{T_d}=0.0411\pm 0.0028$ for a proton, $f^n_{T_u}=0.0189\pm 0.0014$ and $f^n_{T_d}=0.0451\pm 0.0027$ for a neutron \cite{DDupdate}, and  $f^{p,n}_{T_s}=0.043\pm 0.011$ for both proton and neutron \cite{DDstrange}.

We also note that the singlet mediator couplings, $A_S$ and $A_\Phi$, are given from Appendix B, as follows,
\bea
A_S &=&- \frac{1}{\left( 1 + \frac{n^{2} v^2_S}{4 v^2_\Phi} \right)} \frac{q^2}{v^2_S}+\frac{2M^2_A}{v^2_S}\left( \frac{1}{\left( 1 + \frac{n^{2} v^2_S}{4 v^2_\Phi} \right)}-1\right),  \\
A_\Phi &=& - \frac{1}{\left( 1 + \frac{n^{2} v^2_S}{4 v^2_\Phi} \right)} \bigg(\frac{n^2 v^3_S}{4v^3_\Phi}\bigg)\frac{q^2}{v^2_S}+\frac{M^2_A}{v_Sv_\Phi}\left( \frac{1}{\left( 1 + \frac{n^{2} v^2_S}{4 v^2_\Phi} \right)}\,\frac{n^2 v^2_S}{2v^2_\Phi}-n\right).
\eea

For instance, for $\theta\equiv \theta_{12}\neq 0$, $\alpha\equiv\theta_{23}\neq 0$ and $\theta_{13}=0$, the effective DM coupling in eq.~(\ref{DMeff}) becomes
\bea
{\tilde\lambda}=\bigg(\frac{1}{m^2_{h_1}} -\frac{1}{m^2_{h_2}}  \bigg) \Big(c_\alpha A_S-s_\alpha A_\Phi\Big)s_\theta c_\theta.
\eea
Taking the limit of $v_\Phi\gg v_S$ and $m_{h_3}\gg m_{h_2}$, we find that 
\bea
A_S &\simeq& -\frac{q^2}{v^2_S} -\frac{M^2_A}{2v^2_S} \bigg(\frac{n v_S}{v_\Phi}\bigg)^2, \\
A_\Phi &\simeq & -  \bigg(\frac{ v_S}{v_\Phi}\bigg)^3\frac{n^2 q^2}{4v^2_S} -\frac{M^2_A}{v^2_S}\bigg(\frac{n v_S}{v_\Phi}\bigg),
\eea
and
\bea
\sin(2\alpha)\simeq -\frac{\lambda_{\Phi S}}{c_\theta\, \lambda_\Phi}\bigg(\frac{v_S}{v_\Phi}\bigg).
\eea
Thus, the effective DM coupling becomes approximated to
\bea
{\tilde\lambda}\simeq s_\theta c_\theta\bigg(\frac{1}{m^2_{h_1}} -\frac{1}{m^2_{h_2}}  \bigg)\bigg[ -\frac{q^2}{v^2_S} -\frac{M^2_A}{2v^2_S}   \bigg(\frac{v_S}{v_\Phi}\bigg)^2\bigg(n^2+ \frac{n\lambda_{\Phi S}}{c_\theta\, \lambda_\Phi}\bigg) \bigg].
\eea
As a result, the effective DM coupling is small for direct detection experiments, independent of $\theta$, because the first term is suppressed for a small momentum transfer and the second term is suppressed for $v_\Phi\gg v_S$ and $|\lambda_{\Phi S}|\sim \lambda_\Phi$.

\begin{figure}[!t]
\begin{center}
 \includegraphics[width=0.48\textwidth,clip]{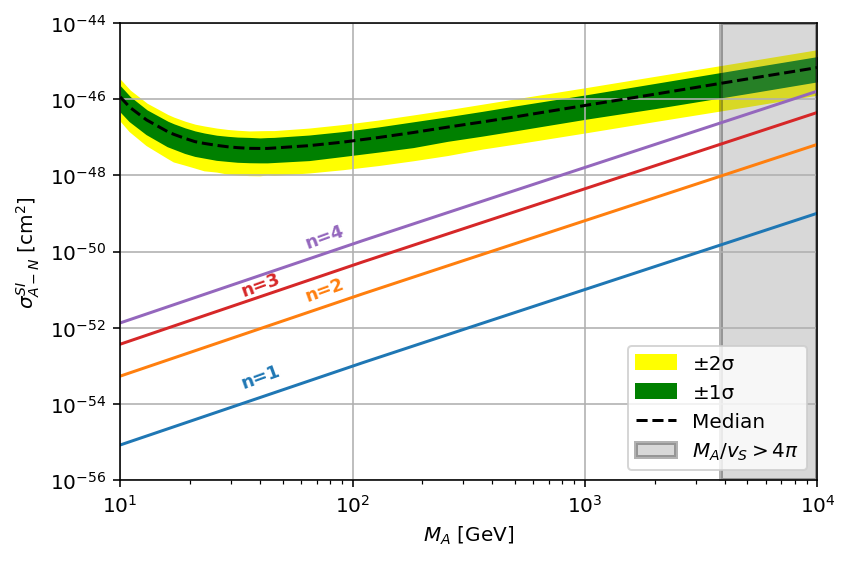} \,\,\,
  \includegraphics[width=0.48\textwidth,clip]{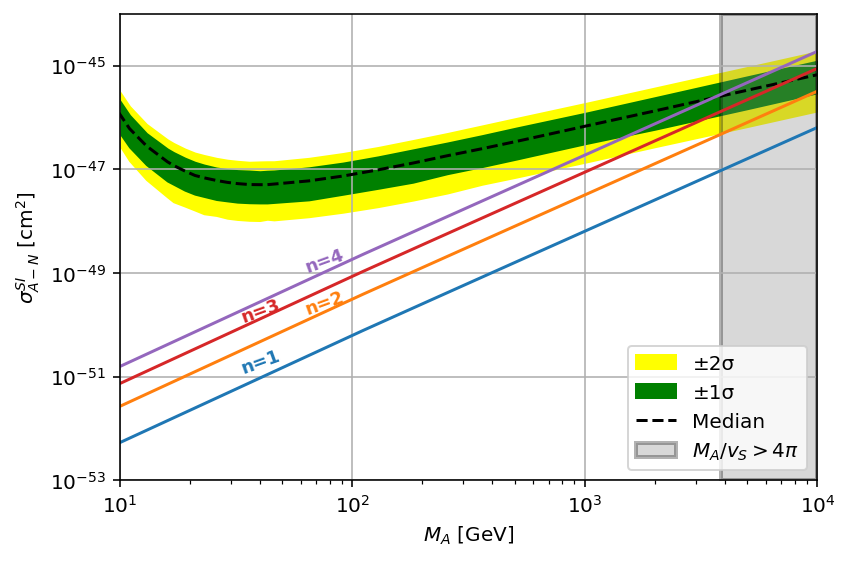} 
 \end{center}
\caption{DM-nucleus scattering cross section as a function of the DM mass $M_A[{\rm GeV}]$ for varying $n$ and $\theta_{13}=0(0.01)$ on the left(right) plots. The green and yellow strips are the bounds from the LZ experiment at $1\sigma$ and $2\sigma$ levels, respectively \cite{LZ}. We also indicated the regions with $M_A/v_S>4\pi$ in gray in which PSDM is not a pNGB. }
\label{fig:DD1}
\end{figure}

In Fig. 1, we show the cross section for  spin-independent DM-nucleus scattering as a function of the DM mass. We set $v_S = 300 \text{ GeV}, v_\Phi = 10 \text{ TeV}, M_A = [10 \text{ GeV}, 10 \text{ TeV}], m_{h_2} = 300 \text{ GeV}, m_{h_3} = 10 \text{ TeV}$, and $\theta_{12}\equiv\theta = 0.1,  \theta_{23}\equiv\alpha = 0.01$, and $\theta_{13}=0 (0.01)$, on the left(right) plots. Accordingly, we calculated the quartic couplings, based on the equations in Appendix A, which are summarized in Table \ref{tab:lambda_values_1}. 
The gray regions are disfavored by the condition that pseudo-scalar dark matter is a pNGB such that the polar basis for scalar fields is appropriate, namely, $M_A/v_S<4\pi$. 

The larger $n$, the larger the direct detection cross sections, as shown in both plots of Fig.~1. 
As in the right plot of Fig. 1, we also find that the direct detection cross section becomes larger as a small mixing angle $\theta_{13}$ is switched on, and the region with DM masses beyond $M_A\sim 1\,{\rm TeV}$ is being excluded by the LZ experiment \cite{LZ}. On the other hand, as compared to the minimal model for PSDM proposed in Ref.~\cite{PSDM}, we keep the same effective mass term violating the $U(1)$ global symmetry for $S$, but the resultant direct detection cross section depends not only on the origin of the effective mass term for $S$ (i.e. $\Phi^n S^2$) but also on the Higgs mixing with an extra Higgs-like scalar coming from $\Phi$.

Moreover, we find that the values of the Higgs quartic coupling $\lambda_H$ in Table \ref{tab:lambda_values_1} are slightly larger than the one with vanishing Higgs mixing angles, namely, $\lambda_H=0.129$, so the vacuum stability can be ensured until a higher scale than in the SM. A nonzero mixing between the Higgs and $\rho_\Phi$ makes the Higgs quartic coupling even larger, as shown in the case with $\theta_{13}=0.01$. The values of the quartic couplings in the model depend on the choice of $\theta_{13}$ in Table \ref{tab:lambda_values_1}, but a reasonable choice of the set of the quartic couplings can be made, being consistent with a small $\theta_{13}$.

We note that for $v_S = 300 \text{ GeV}$ and $v_\Phi = 10 \text{ TeV}$,  the DM masses in the range of $M_A = [10 \text{ GeV}, 10 \text{ TeV}]$ are translated to the values of  $\kappa_n$ and $\Lambda$: $\kappa_1=3.5\times 10^{-7}-0.35$ for $\Lambda=v_\Phi$, $\kappa_2=5.0\times 10^{-7}-0.50$, $\kappa_3=7.1\times 10^{-6}-0.71$ for $\Lambda=10v_\Phi$, and $\kappa_4=1.0\times 10^{-4}-1.0$ for $\Lambda=10v_\Phi$.

\begin{table}[h!]
    \centering
    \begin{tabular}{|c|c|c|c|c|}
        \hline
        $n$ &  $\theta_{13}$  & $\lambda_{HS}$ & $\lambda_{H\Phi}$ & $\lambda_{\Phi S}$ \\ \hline
        1 &  0   & $-0.113$ & 0.0405 & $[-0.331, 0.169]$ \\ \hline
        2 & 0   & $-0.113$ & 0.0405 & $[-0.331,0.668]$ \\ \hline
        3 & 0   & $-0.113$ & 0.0405 & $[-0.331,1.17]$ \\ \hline
        4 & 0  & $-0.113$ & 0.0405 & $[-0.331,1.66]$ \\ \hline
        1 &  0.01  & 0.0327 & $-0.364$ & $[-0.365, 0.135]$ \\ \hline
        2 &  0.01  & 0.0327 & $-0.364$ & $[-0.365, 0.635]$ \\ \hline
        3 &  0.01 & 0.0327 & $-0.364$ & $[-0.365, 1.13]$ \\ \hline
        4 &  0.01 & 0.0327 & $-0.364$ & $[-0.365, 1.63]$ \\ \hline
    \end{tabular}
    \caption{Quartic couplings for varying $n$ in $\Phi^n S^2$, used for Fig. 1. We also took $\lambda_{H}=0.137(0.202)$, $\lambda_\Phi=0.50$, and $\lambda_S=0.551(0.562)$, for $\theta_{13}=0(0.01)$.}
    \label{tab:lambda_values_1}
\end{table}

\begin{figure}[!t]
\begin{center}
 \includegraphics[width=0.48\textwidth,clip]{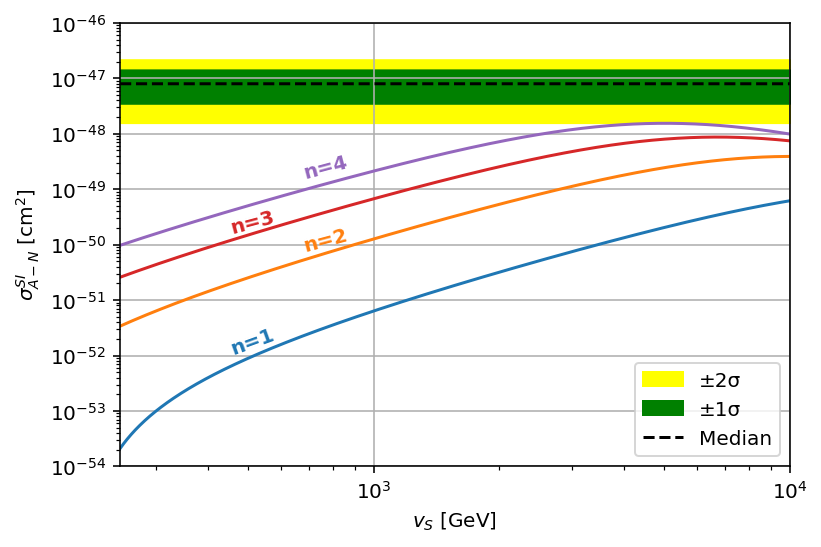} \,\,\,
  \includegraphics[width=0.48\textwidth,clip]{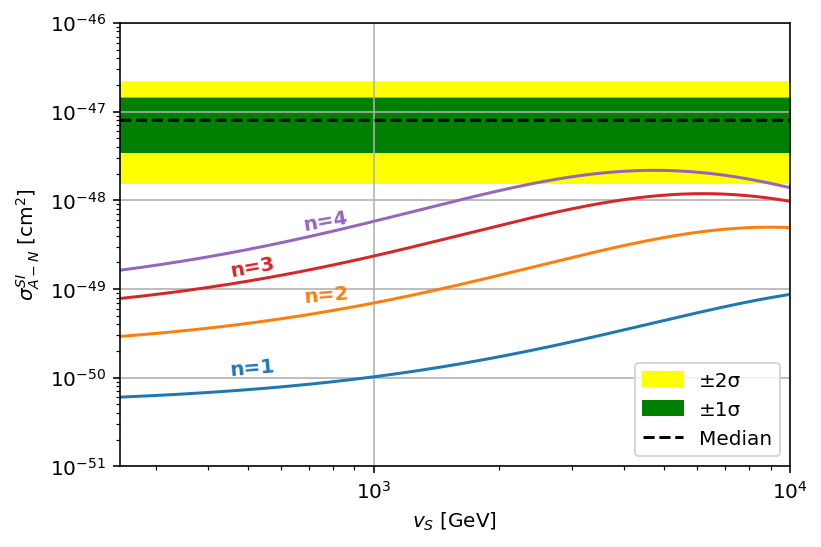}
 \end{center}
\caption{DM-nucleus scattering cross section as a function of $v_S[{\rm GeV}]$ for varying $n$ and $\theta_{13}=0(0.01)$ on the left(right) plots. In all the regions of the parameters space,  $M_A/v_S<4\pi$ is satisfied. }
\label{fig:DD2}
\end{figure}

In Fig. 2, we show the cross section for  spin-independent DM-nucleus scattering as a function of $v_S$. We set $v_S = [v,v_\Phi], v_\Phi = 10 \text{ TeV}, M_A=100\,{\rm GeV},  m_{h_2} = 300 \text{ GeV}, m_{h_3} = 10 \text{ TeV}$ and $\theta = 0.1,  \alpha = 0.01$, and $\theta_{13}=0 (0.01)$, on the left(right) plots. In this case, we also calculated the quartic couplings, based on the equations in Appendix A, which are summarized in Table \ref{tab:lambda_values_2}. In most of the parameter space with $v_S = [v,v_\Phi]$, the model is consistent with direct detection bounds, independent of $n$.
Again we find that we can make a reasonable choice of the set of the quartic couplings  in Table \ref{tab:lambda_values_2} to be consistent with a small $\theta_{13}$.
We note that all the parameter space in Fig.~\ref{fig:DD2} is consistent with $M_A/v_S<4\pi$.

Varying $v_S$ in the range of $[v,v_\Phi]$, we find that the DM mass is not so sensitive to the choice of $v_S$ as shown in eq.~(\ref{Amass}), and the values of $\kappa_n$ necessary to get $M_A=100\,{\rm GeV}$ are determined to be $\kappa_1\lesssim 3.5\times 10^{-5}$ for $\Lambda=v_\Phi$, $\kappa_2\lesssim 5.0\times 10^{-5}$, $\kappa_3\lesssim 7.1\times 10^{-5}$ for $\Lambda=10v_\Phi$, and $\kappa_4\lesssim 1.0\times 10^{-4}$ for $\Lambda=10v_\Phi$.

\begin{table}[h!]
    \centering 
    \begin{tabular}{|c|c|c|c|c|}
        \hline
        $n$ &  $\theta_{13}$  & $\lambda_S$ & $\lambda_{HS}$  & $-\lambda_{\Phi S}$ \\ \hline
        1 &  0 &  $[0.000496, 0.819]$ & $[-0.138,-0.00340] $ &  $ [0.00990, 0.404]$ \\ \hline
        2 &  0 &  $[0.000496,0.819]$ &    $[ -0.138,-0.00340] $ &  $[0.00989, 0.404] $ \\ \hline
        3 &  0 &  $[0.000496,0.819]$  & $[-0.138,-0.00340] $ &  $[0.00989, 0.404] $ \\ \hline
        4 &  0 &  $[0.000496,0.819]$ & $[-0.138,-0.00340] $ &  $[0.00990, 0.404] $ \\ \hline
        1 & 0.01 &  $[0.000506, 0.836]$  & $[0.000980, 0.0398]$ & $[0.0109,0.445]$ \\ \hline
        2 & 0.01 &  $[0.000506, 0.836]$  & $[0.000980, 0.0398]$ & $[0.0109,0.445 ]$ \\ \hline
        3 & 0.01 &  $[0.000506, 0.836]$  & $[0.000980, 0.0398]$ & $[0.0109,0.444]$ \\ \hline
        4 & 0.01 &  $[0.000506, 0.836]$  & $[0.000980, 0.0398]$ & $[0.0109,0.444]$ \\ \hline
    \end{tabular}
    \caption{Quartic couplings for varying $n$ in $\Phi^n S^2$, used for Fig. 2. We also took $\lambda_H=0.137(0.202)$, $\lambda_\Phi=0.500$, and  $\lambda_{H\Phi}=0.0405(-0.364)$, for $\theta_{13}=0(0.01)$.}
    \label{tab:lambda_values_2}
\end{table}

\subsection{Dark matter relic density}

PSDM can self-annihilate into the SM fermions or massive gauge bosons by $AA\to {\bar f}f, WW,ZZ$ through the Higgs-portal couplings and into  a pair of Higgs-like scalars by $AA\to h_i h_j$ with $i,j=1,2,3$.
Then, the relic density for PSDM is governed by the following Boltzmann equation, 
\bea
{\dot n}_A+ 3H n_A=-\langle(\sigma v)_{\rm tot}\rangle \Big(n^2_A-(n^{\rm eq}_A)^2 \Big),
\eea
where $\langle(\sigma v)_{\rm tot}\rangle$ is the averaged total annihilation cross section,  with
\bea
(\sigma v)_{\rm tot}=2(\sigma v)_{AA\to {\bar f}f}+2(\sigma v)_{AA\to WW}+ 2(\sigma v)_{AA\to ZZ}+2\sum_{i\leq j}(\sigma v)_{AA\to h_i h_j}.
\eea
Here, the factor $2$ in front of the cross sections is due to the fact that the real scalar PSDM is annihilated in a pair per annihilation.
Then, the relic density for PSDM can be determined by
\bea
\Omega_A h^2=0.2745 \bigg(\frac{Y_A}{10^{-11}} \bigg)\bigg(\frac{M_A}{100\,{\rm GeV}} \bigg)
\eea
where $Y_A=n_A/s$ is the DM abundance at present.

\begin{figure}[!t]
\begin{center}
 \includegraphics[width=0.40\textwidth,clip]{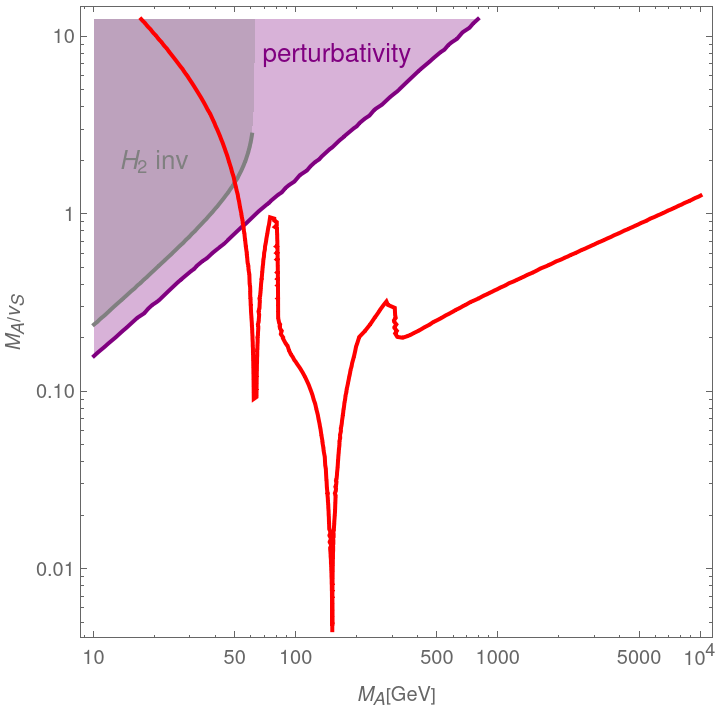} \,\,\,
  \includegraphics[width=0.40\textwidth,clip]{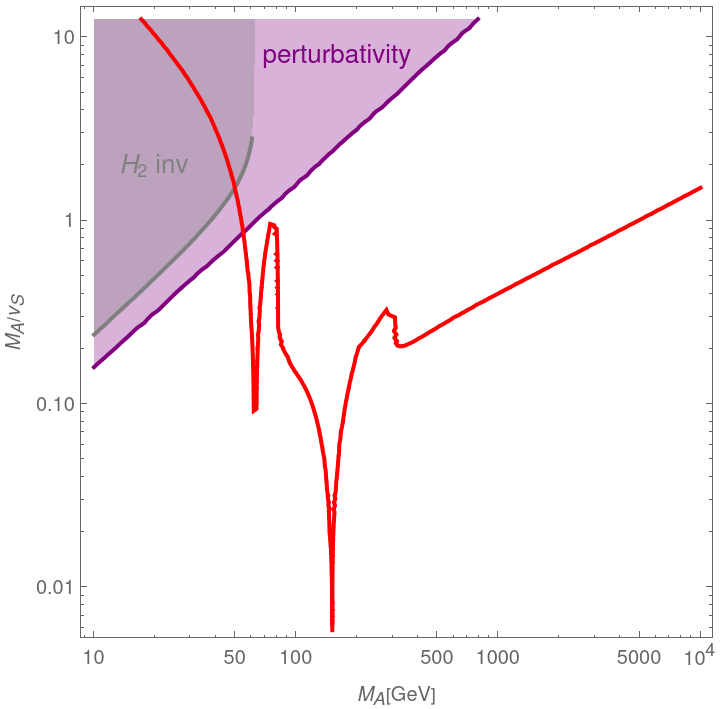}
 \end{center}
\caption{Parameter space in $M_A [{\rm GeV}]$ vs $M_A/v_S$ for the correct relic density with  $\theta_{13}=0$ in red. The entire parameter space shown in the plots is consistent with the current direct detection bounds from LZ. We took $n=1, 4$ in $\Phi^nS^2$ on the left and right plots, respectively. The parameter choice is the same as in the left plot of Fig.~\ref{fig:DD1}. We overlaid the region excluded by the bounds on the Higgs invisible decays in gray and the region disfavored by the perturbativity bound $\lambda_S<4\pi$ in purple. }
\label{fig:relic1}
\end{figure}

In Fig.~\ref{fig:relic1}, we depict the parameter space in dark matter mass $M_A$ vs the effective DM coupling $M_A/v_S$, explaining the correct relic density in red lines and satisfying the direct detection bounds from LZ. We chose $\theta_{13}=0$, and $n=1,4$ in $\Phi^n S^2$ on the left and right plots, respectively. We indicated the region disfavored by the perturbativity bound, $\lambda_S<4\pi$, in purple and the bound the Higgs invisible decay rules out the gray regions. We chose the parameters as in the left plot of Fig.~\ref{fig:DD1}, in particular, $m_{h_2}=300\,{\rm GeV}$. But, if we take a smaller $m_{h_2}$, $\lambda_S$ can be smaller for the same value of $v_S$, so  the perturbativity bound in purple gets weaker and there is a more parameter space for PSDM that is consistent with indirect detection, because $AA\to h_2 h_2$ starts being important above a smaller mass of PSDM. 

We also remark that the resonance annihilation channels allow for a smaller effective DM coupling for $M_A\sim m_{h_1}/2$ or  $m_{h_2}/2$, and the new annihilation channel, $AA\to h_2 h_2$, with $h_2$ being the lighter singlet-like scalar, also makes a smaller value of the effective DM coupling, $M_A/v_S$, for $M_A>m_{h_2}$, to be compatible with the correct relic density. We find that similar results can be obtained for other values of $n$, as far as the corresponding mixing term between the pseudo-scalar fields, $\Phi^n S^2$, is constrained to give rise to an appropriate mass for pseudo-scalar dark matter.

\subsection{Indirect detection}

Indirect detection bounds from Fermi-LAT \cite{Fermilat} are relevant for PSDM annihilating into the SM particles, $WW$,  $b{\bar b}$ or $h_1h_1$, etc, in our model, and the model can be constrained further by the future LSST \cite{LSST} and CTA  experiments \cite{CTA}. The Fermi-LAT dSphs set the limit on $AA\to b{\bar b}$  to be below the thermal cross section for $M_A\lesssim 200\,{\rm GeV}$ \cite{Fermilat}, so the region with a relatively light PSDM where $AA\to b{\bar b}$ dominates would be in tension with indirect detection. 

On the other hand, when $AA\to WW, ZZ$ or $h_1h_1$ open up, they are dominant and the ratios of the annihilation cross sections for $AA\to WW,ZZ,h_1h_1$ become $WW:ZZ:h_1h_1\simeq 2:1:1$, in the limit of small Higgs mixing angles, due to the Goldstone equivalence theorem.  The bound on $AA\to WW$ is not as strong as the one on  $AA\to b{\bar b}$ \cite{Fermilat}, so the region with a heavier PSDM with $M_V\gtrsim M_W$ is consistent with indirect detection.

\subsection{Higgs data and collider bounds}

For $M_A<m_{h_1}/2$, the SM-like Higgs can decay invisibly into a pair of PSDM particles, with the corresponding decay rate given by
\bea
\Gamma(h_1\to AA)&=&\frac{v^2_S}{32\pi m_{h_1}}\, \bigg|R_{21} A_S + R_{31} A_\Phi \bigg|^2_{q^2=m^2_{h_1}} \sqrt{1-\frac{4M^2_A}{m^2_{h_1}}}.
\eea
Then, for $\theta_{13}=0$, we get
\bea
\Gamma(h_1\to AA)&=&\frac{m^3_{h_1}\sin^2\theta}{32\pi v^2_S}\,\bigg(1+\frac{n^2 v^2_S}{4v^2_\Phi}\bigg)^{-2}\bigg[ \cos\alpha\bigg(1+\frac{M_A^2 n^2 v^2_S}{2m^2_{h_1}v^2_\Phi}\bigg) \nonumber \\
&&-\frac{n v_S}{v_\Phi}\,\sin\alpha\bigg\{ \frac{M_A^2}{m^2_{h_1}} \bigg(1+\frac{n(n-2) v^2_S}{4v^2_\Phi} \bigg)+\frac{n v^2_S}{4 v^2_\Phi}\bigg\} \bigg]^2\sqrt{1-\frac{4M^2_A}{m^2_{h_1}}}.
\eea
For $v_\Phi\gg v_S$, the above result gets simplified to
\bea
\Gamma(h_1\to AA})\simeq\frac{m^3_{h_1}\sin^2\theta\cos^2\alpha}{32\pi v^2_S}\,\sqrt{1-\frac{4M^2_A}{m^2_{h_1}},
\eea
independent of $n$.
The bounds on the branching ratio of the Higgs invisible decay are given by ${\rm BR}(h_1\to {\rm invisible})<0.107$ in ATLAS and  ${\rm BR}(h_1\to {\rm invisible})<0.15$ in CMS at 95\% C.L.  \cite{Hinv}, so the region with $M_A<m_{h_1}/2$ is constrained by the bounds on the Higgs invisible decays in our model. Moreover, nonzero Higgs mixing angles would reduce the Higgs signal strength universally, so $\sin^2\theta$ can be constrained to be smaller than about 0.1 by the Higgs data.

\section{Conclusions}
\label{conclusion}

We presented a new model for pseudo-scalar dark matter in the extension of the SM with an extra gauged $U(1)_X$ symmetry. 
We found that the $U(1)_X$ symmetry gives rise to the $Z_2$ symmetry as an accidental symmetry at the renormalizable level, and the stability of pseudo-scalar dark matter is ensured in the absence of $U(1)_X$-charged dark fermions in the model. After the $U(1)_X$ symmetry is broken by the VEVs of the singlet scalar fields, $v_\Phi$ and $v_S$, the $Z_2$ symmetry remains unbroken, so the combination of pseudo-scalar fields in the dark sector appears as a stable pNGB dark matter receiving the mass $M_A$ from the $U(1)_X$ invariant mixing potential. 

In the presence of the hierarchy, $M_A\ll  v_S \ll  v_\Phi$, the $U(1)_X$ symmetry is broken spontaneously, dominantly by $v_\Phi$, so the extra gauge boson is decoupled, whereas pseudo-scalar dark matter appears light as a pNGB having both derivative and non-derivative interactions in the polar basis for scalar fields. 
In this case, the direct detection cross section for pseudo-scalar dark matter  gets suppressed by the momentum transfer between DM and the nucleus, being insensitive to the Higgs-portal couplings, but the correct relic density can be explained by the DM annihilations into the SM particles  or into a pair of light Higgs-like scalars.   Thus, the Higgs-portal couplings are constrained mostly by the Higgs invisible decays and the Higgs data.

Even for a mildly decoupled new scalar field is or extra gauge boson, the resultant direct detection cross section for PSDM remains suppressed for weak-scale DM masses, being compatible with the current bounds from direct detection, but the case for a relatively heavy PSDM depends not only on the origin of the effective mass term for $S$ (i.e. $\Phi^n S^2$) but also on the Higgs mixing with an extra Higgs-like scalar, as compared to the minimal model in Ref.~\cite{PSDM}.

%%%%%%%%%%%%%%%%%%%%%%%%%%%%%%%%%%%%%%%%%%%%%%%%%%%%%%%%%%%%%%%%%%%
\section*{Acknowledgements}
%%%%%%%%%%%%%%%%%%%%%%%%%%%%%%%%%%%%%%%%%%%%%%%%%%%%%%%%%%%%%%%%%%%

This work is supported in part by Basic Science Research Program through the National
Research Foundation of Korea (NRF) funded by the Ministry of Education, Science and
Technology (NRF-2022R1A2C2003567(JHK and HML) and RS-2024-00341419(JKK)). 
This research was supported by the Chung-Ang University Graduate Research Scholarship in 2025.

%%%%%%%%%%%%%%%%%%%%%%%%%%%%%%%%%%%%%%%%%%%%%%%%%%%%%%%%%%%%%%%%%%%
%%%%%%%%%%%%%%%%%%%%%%%%%%%%%%%%%%%%
\appendix 
%%%%%%%%%%%%%%%%%%%%%%%%%%%%%%%%%%%%
\section{Bounds on quartic couplings}\label{quartic}

We can trade off the quartic couplings to scalar masses and Higgs mixing angles, as follows,
\bea
\lambda_H &=& \frac{1}{2v^2}\bigg(\sum_{i=1}^3 m_{h_i}^2 R_{i1}^2  \bigg), \\
\lambda_S &=& \frac{1}{2v^2_S}\bigg(\sum_{i=1}^3 m_{h_i}^2 R_{i2}^2  \bigg), \\
\lambda_\Phi &=&  \frac{1}{2v^2_\Phi}\left(\sum_{i=1}^3 m_{h_i}^2 R_{i3}^2 +  \frac{n(n-2) v^2_S}{4v_\Phi^2 \left( 1 + \frac{n^{2} v^2_S}{4 v^2_\Phi} \right)} M^2_A \right), \\
\lambda_{HS} &=&\frac{1}{v v_S} \bigg(\sum_{i=1}^3 m_{h_i}^2 R_{i1}R_{i2}  \bigg),  \\
\lambda_{H\Phi} &=&\frac{1}{v v_\Phi} \bigg(\sum_{i=1}^3 m_{h_i}^2 R_{i1}R_{i3}  \bigg),  \label{the13} \\
\lambda_{\Phi S}&=& \frac{1}{v_S v_\Phi} \left(\sum_{i=1}^3 m_{h_i}^2 R_{i2}R_{i3} +  \frac{n v_S}{2 v_\Phi \left( 1 + \frac{n^{2} v^2_S}{4 v^2_\Phi} \right)}  M^2_A \right).
\eea
Then, we can constrain the general masses and mixing angles in our model from the perturbativity and unitarity bounds on the quartic couplings \cite{ligong}.

For instance, for $\theta\equiv \theta_{12}\neq 0$, $\alpha\equiv\theta_{23}\neq 0$ and $\theta_{13}=0$, the mixing quartic couplings are related to the mixing angles by
\bea
\lambda_{HS} &=&\frac{1}{v v_S} \,  \Big(m^2_{h_1} -c^2_\alpha  m^2_{h_2}- s^2_\alpha m^2_{h_3}\Big)c_\theta s_\theta, \\
\lambda_{H\Phi} &=&\frac{1}{v v_\Phi} \Big(m^2_{h_3}-m^2_{h_2} \Big)s_\theta c_\alpha s_\alpha,  \\
\lambda_{\Phi S} &=&  \frac{1}{v_S v_\Phi} \left(\Big(m^2_{h_2}-m^2_{h_3}\Big)c_\theta c_\alpha s_\alpha+  \frac{n v_S}{2 v_\Phi \left( 1 + \frac{n^{2} v^2_S}{4 v^2_\Phi} \right)}  M^2_A \right).
\eea
In this case, we note that $\lambda_{H\Phi}$ can be nonzero even if $\theta_{13}= 0$.
Taking $v_\Phi\gg v_S, v, M_A$ and $m_{h_3}\simeq \sqrt{2\lambda_\Phi}  \,v_\Phi\gg m_{h_2}\simeq \sqrt{2\lambda_S}\, v_S$, we can determine the mixing angle $\alpha$ by
\bea
\sin(2\alpha)\simeq -\frac{2\lambda_{\Phi S}v_S v_\Phi}{\cos\theta\,  m^2_{h_3}}\simeq -\frac{\lambda_{\Phi S} }{\cos\theta\,\lambda_\Phi}\bigg(\frac{v_S}{ v_\Phi}\bigg)
\eea
where we used $m^2_{h_3}\simeq 2\lambda_\Phi v^2_\Phi$, so $|\alpha|\ll 1$ for $|\lambda_{\Phi S}|\sim \lambda_\Phi$.  The results are used in the text.

\section{Pseudo-scalar couplings}\label{psdmcouplings}

In the absence of the mixings between CP-even scalars, the Feynman rules for the triple vertex interactions of the pseudo-scalar $A$ are
{\footnotesize
\bea
&&{[}A(p_1)A(p_2)\rho_S]: iv_S \bigg[ - \frac{1}{\Big( 1 + \frac{n^{2} v^2_S}{4 v^2_\Phi} \Big)} \frac{q^2}{v^2_S}+\frac{2M^2_A}{v^2_S}\bigg( \frac{1}{\Big( 1 + \frac{n^{2} v^2_S}{4 v^2_\Phi} \Big)}-1\bigg) \bigg]\equiv i v_S A_S, \label{As}\\
&&{[}A(p_1)A(p_2)\rho_\Phi]:  iv_S  \bigg[ - \frac{1}{\Big( 1 + \frac{n^{2} v^2_S}{4 v^2_\Phi} \Big)} \bigg(\frac{n^2 v^3_S}{4v^3_\Phi}\bigg)\frac{q^2}{v^2_S}+\frac{M^2_A}{v_Sv_\Phi}\bigg( \frac{1}{\Big( 1 + \frac{n^{2} v^2_S}{4 v^2_\Phi} \Big)}\,\frac{n^2 v^2_S}{2v^2_\Phi}-n\bigg) \bigg] \equiv i v_S A_\Phi. \label{Ap}
\eea
}
Here, $p_1$ is the incoming momentum, $p_2$ is the outgoing momentum, and $q=p_1-p_2$.

Focusing on the interaction for $A(p_1)A(p_2)\rho_S$, the terms proportional to $M^2_A$ are not cancelled for a finite value of $v_\Phi$ in eq.~(\ref{As}), unlike the case where the global $U(1)$ breaking mass parameter  for $S$ is constant \cite{PSDM}.  But, in the limit of $v_\Phi\gg v_S$, those terms become small, so the direct detection cross sections for the Higgs mixing with $\rho_S$ get suppressed by the momentum transfer $q^2$, even for a sizable mixing between the SM Higgs and $\rho_S$. On the other hand, the interaction for $A(p_1)A(p_2)\rho_\Phi$ can be made decoupled from direct detection, as far as $\rho_\Phi$ is decoupled from $h$ and $\rho_S$.

In general, including the mixings between CP-even scalars as in eq.~(\ref{mixing}), we find the triple interactions between the mass eigenstates, $(h_1, h_2, h_3)$ and the pseudo-scalar as
\bea
{[}A(p_1)A(p_2)h_i]:&& i v_S \Big( R_{2i}A_S+R_{3i} A_\Phi \Big), \quad i=1,2,3.
\eea

For instance, for $\theta\equiv \theta_{12}\neq 0$, $\alpha\equiv\theta_{23}\neq 0$ and $\theta_{13}=0$, the triple interactions between the CP-even states and the pseudo-scalar become
\bea
{[}A(p_1)A(p_2)h_1]:&& iv_S \Big( -c_\alpha A_S+s_\alpha A_\Phi\Big)s_\theta,  \\
{[}A(p_1)A(p_2)h_2]:&& iv_S \Big( c_\alpha A_S-s_\alpha A_\Phi\Big)c_\theta,  \\
{[}A(p_1)A(p_2)h_3]:&& iv_S \Big( s_\alpha A_S+c_\alpha A_\Phi\Big).
\eea

\section{PSDM annihilation cross sections}\label{ann}

The annihilation channels for PSDM are composed of $AA\to f{\bar f}, WW, ZZ$ and $h_i h_j$ with $i,j=1,2,3$.
In the non-relativistic limit for PSDM, the corresponding annihilation cross sections are given by
\bea
(\sigma v)_{AA\to f{\bar f}}=\frac{1}{32\pi M^2_A}\,|M_{AA\to f{\bar f}}|^2\sqrt{1-\frac{m^2_f}{M^2_A}},
\eea 
with
\bea
|M_{AA\to f{\bar f}}|^2= 4N_c M^2_A\left| \sum_{i=1}^3\frac{\lambda_{h_i AA}(4M^2_A) \cdot y_{h_i f{\bar f}}}{4M^2_A-m^2_{h_i}} \right|^2 \bigg(1-\frac{m^2_f}{M^2_A}\bigg),
\eea
and
\bea
(\sigma v)_{AA\to VV}=\frac{\delta_V}{32\pi M^2_A}\,|M_{AA\to VV}|^2\sqrt{1-\frac{M^2_V}{M^2_A}}, \quad V=W,Z,
\eea
with $\delta_V=1,\frac{1}{2}$ for $V=W, Z$, and
\bea
|M_{AA\to VV}|^2=\left| \sum_{i=1}^3\frac{\lambda_{h_i AA}(4M^2_A) \cdot g_{h_i VV}}{4M^2_A-m^2_{h_i}} \right|^2\bigg(3-\frac{4M^2_A}{M^2_V}+\frac{4M^4_A}{M^4_V}\bigg),
\eea
and
\bea
(\sigma v)_{AA\to h_i h_j}=\frac{1-\frac{1}{2}\delta_{ij}}{32\pi M^2_A}\,|M_{AA\to h_i h_j}|^2\sqrt{1-\frac{(m_{h_i}+m_{h_j})^2}{4M^2_A}}\sqrt{1-\frac{(m_{h_i}-m_{h_j})^2}{4M^2_A}},
\eea
with
\bea
|M_{AA\to h_i h_j}|^2&=&\bigg| \lambda_{h_i h_j AA}(4M^2_A)+\lambda_{h_iAA}(m^2_{h_i})\cdot\lambda_{h_jAA}(m^2_{h_j})\bigg(\frac{1}{t-M^2_A}+\frac{1}{u-M^2_A}\bigg) \nonumber \\
&&+\sum_{k=1}^3\lambda_{h_kAA}(4M^2_A)\cdot\lambda_{h_i h_j h_k}\,\cdot\frac{1}{4M^2_A-m^2_{h_k}} \bigg|^2,
\eea
and $t\simeq \frac{1}{2}(m^2_{h_i}+m^2_{h_j}-2M^2_A)\simeq u$.
Here, the effective interactions vertices for PSDM with fermions, massive gauge bosons or Higgs-like scalars are denoted by
\bea
y_{h_i f{\bar f}} &=& -\frac{m_f}{v}\, R_{1i}, \\
g_{h_i VV}&=& \frac{2M^2_V}{v}\, R_{1i}, \\
\lambda_{h_i AA}(q^2) &=& v_S (R_{2i}A_S(q^2) + R_{3i} A_\Phi (q^2)), \\
\lambda_{h_i h_j AA}(q^2)&=& \frac{M^2_A}{v^2_S} \bigg[2R_{2i} \Big(R_{2j}+ nR_{3j} \frac{v_S}{v_\phi}\Big)+nR_{3i}\frac{v_S}{v_\Phi} \bigg(2R_{2j}+(n-1) R_{3j} \frac{v_S}{v_\Phi}\bigg) \bigg] \nonumber \\
&&+\frac{1}{4v^2_S}(q^2-2M^2_A) \bigg(4R_{2i} R_{2j}+ n^2 R_{3i}R_{3j} \frac{v^4_S}{v^4_\Phi} \bigg) \bigg(1+\frac{n^2 v^2_S}{4v^2_\Phi}\bigg)^{-1},
\eea 
and $\lambda_{h_i h_j h_k}$ are the cubic couplings between $h_i$, $h_j$ and $h_k$.
We note that $A_S, A_\Phi$ are defined in eqs.~(\ref{As}) and (\ref{Ap}).
In the limit of $M_A\gtrsim M_V, m_{h_1}$ and small Higgs mixing angles, the ratios of the annihilation cross sections for $AA\to WW,ZZ,h_1h_1$ become $WW:ZZ:h_1h_1\simeq 2:1:1$, due to the Goldstone equivalence theorem. 

For instance, for $\theta\equiv \theta_{12}\neq 0$, $\alpha\equiv\theta_{23}\neq 0$ and $\theta_{13}=0$, the nonzero effective interactions for PSDM are given by
\bea
-y_{h_1 f{\bar f}}/(m_f/v) &=& g_{h_1VV}/(2M^2_V/v)=\cos\theta, \\
 -y_{h_2 f{\bar f}}/(m_f/v) &=& g_{h_2VV}/(2M^2_V/v)=\sin\theta,  \\
 \lambda_{h_1 AA} &=& v_S \Big(-\cos\alpha\, A_S+\sin\alpha A_\Phi \Big)\sin\theta, \\
\lambda_{h_2 AA} &=& v_S \Big(\cos\alpha\, A_S-\sin\alpha A_\Phi \Big)\cos\theta, \\
\lambda_{h_3 AA} &=& v_S \Big(\sin\alpha\, A_S+\cos\alpha A_\Phi \Big),
\eea
and
\bea
\lambda_{h_1 h_1 AA} &=&\frac{M^2_A}{v^2_S}\, s^2_\theta \bigg[2 c_\alpha \Big(c_\alpha- n s_\alpha \frac{v_S}{v_\phi}\Big)+n s_\alpha\frac{v_S}{v_\Phi} \bigg(-2c_\alpha+(n-1) s_\alpha \frac{v_S}{v_\Phi}\bigg) \bigg] \nonumber \\
&&+\frac{1}{4v^2_S}(q^2-2M^2_A) \, s^2_\theta \bigg(4c^2_\alpha+ n^2 s^2_\alpha \frac{v^4_S}{v^4_\Phi} \bigg) \bigg(1+\frac{n^2 v^2_S}{4v^2_\Phi}\bigg)^{-1} \nonumber \\
&\simeq &\bigg[\frac{2M^2_A}{v^2_S}\, + \frac{1}{v^2_S}(q^2-2M^2_A)\bigg] s^2_\theta c^2_\alpha, \\
\lambda_{h_2 h_2 AA} &=&\frac{M^2_A}{v^2_S}\, c^2_\theta \bigg[2 c_\alpha \Big(c_\alpha- n s_\alpha \frac{v_S}{v_\phi}\Big)+n s_\alpha\frac{v_S}{v_\Phi} \bigg(2c_\alpha-(n-1) s_\alpha \frac{v_S}{v_\Phi}\bigg) \bigg] \nonumber \\
&&+\frac{1}{4v^2_S}(q^2-2M^2_A) \, c^2_\theta \bigg(4c^2_\alpha+ n^2 s^2_\alpha \frac{v^4_S}{v^4_\Phi} \bigg) \bigg(1+\frac{n^2 v^2_S}{4v^2_\Phi}\bigg)^{-1} \nonumber \\
&\simeq & \bigg[\frac{2M^2_A}{v^2_S} +\frac{1}{v^2_S}(q^2-2M^2_A)\bigg] c^2_\theta c^2_\alpha, 
\eea
\bea
\lambda_{h_3 h_3 AA} &=&\frac{M^2_A}{v^2_S}\,  \bigg[2 s_\alpha \Big(s_\alpha+ n c_\alpha \frac{v_S}{v_\phi}\Big)+n c_\alpha\frac{v_S}{v_\Phi} \bigg(2s_\alpha+(n-1) c_\alpha \frac{v_S}{v_\Phi}\bigg) \bigg] \nonumber \\
&&+\frac{1}{4v^2_S}(q^2-2M^2_A) \,  \bigg(4s^2_\alpha+ n^2 c^2_\alpha \frac{v^4_S}{v^4_\Phi} \bigg) \bigg(1+\frac{n^2 v^2_S}{4v^2_\Phi}\bigg)^{-1} \nonumber \\
&\simeq& \bigg[\frac{2M^2_A}{v^2_S} +\frac{1}{v^2_S}(q^2-2M^2_A)\bigg]s^2_\alpha,   \\
\lambda_{h_1 h_2 AA} &=&\frac{M^2_A}{v^2_S}\,s_\theta c_\theta  \bigg[-2c_\alpha \Big(c_\alpha- n  s_\alpha \frac{v_S}{v_\phi}\Big)+n  s_\alpha\frac{v_S}{v_\Phi} \bigg(2 c_\alpha-(n-1) s_\alpha  \frac{v_S}{v_\Phi}\bigg) \bigg] \nonumber \\
&&-\frac{1}{4v^2_S}(q^2-2M^2_A) \, s_\theta c_\theta  \bigg(4  c^2_\alpha+ n^2 s^2_\alpha \frac{v^4_S}{v^4_\Phi} \bigg) \bigg(1+\frac{n^2 v^2_S}{4v^2_\Phi}\bigg)^{-1} \nonumber \\
&\simeq &  -\bigg[\frac{2M^2_A}{v^2_S} +\frac{1}{v^2_S}(q^2-2M^2_A)\bigg]s_\theta c_\theta c^2_\alpha , 
\eea
\bea
\lambda_{h_2 h_3 AA} &=&\frac{M^2_A}{v^2_S}\,c_\theta c_\alpha  \bigg[2\Big(s_\alpha- n  c_\alpha \frac{v_S}{v_\phi}\Big)-n  \frac{v_S}{v_\Phi} \bigg(2 s_\alpha+(n-1) c_\alpha  \frac{v_S}{v_\Phi}\bigg) \bigg] \nonumber \\
&&-\frac{1}{4v^2_S}(q^2-2M^2_A) \, c_\theta c_\alpha  \bigg(4  s_\alpha- n^2 s_\alpha \frac{v^4_S}{v^4_\Phi} \bigg) \bigg(1+\frac{n^2 v^2_S}{4v^2_\Phi}\bigg)^{-1} \nonumber \\
&\simeq &  \bigg[\frac{2M^2_A}{v^2_S} +\frac{1}{v^2_S}(q^2-2M^2_A)\bigg] c_\theta c_\alpha s_\alpha,  \\
\lambda_{h_1 h_3 AA} &=&\frac{M^2_A}{v^2_S}\,s_\theta  \bigg[-2c_\alpha \Big(s_\alpha+ n  c_\alpha \frac{v_S}{v_\phi}\Big)+n  s_\alpha\frac{v_S}{v_\Phi} \bigg(2 s_\alpha+(n-1) c_\alpha  \frac{v_S}{v_\Phi}\bigg) \bigg] \nonumber \\
&&-\frac{1}{4v^2_S}(q^2-2M^2_A) \, s_\theta s_\alpha c_\alpha  \bigg(-4 + n^2  \frac{v^4_S}{v^4_\Phi} \bigg) \bigg(1+\frac{n^2 v^2_S}{4v^2_\Phi}\bigg)^{-1} \nonumber \\
&\simeq &  -\bigg[\frac{2M^2_A}{v^2_S} +\frac{1}{v^2_S}(q^2-2M^2_A)\bigg] s_\theta c_\alpha s_\alpha.
\eea
Here, we took $v_\Phi \gg v_S$ in the second equality of each equation in the above, so the effective interactions for PSDM are further simplified. Moreover, we only have to consider the annihilation channels, $AA\to h_1/h_2\to f{\bar f}, WW, ZZ$ and $AA\to h_1 h_1, h_2 h_2, h_1 h_2$, for $m_{h_3}\gtrsim M_A> m_{h_1}, m_{h_2}$.

\medskip

%%%%%%%%%%%%%%%%%%%%%%%%%%%%%%%%%%%%%%%%%%%%%%%%%%%%%%%%%%%%%%%%%%%
%\clearpage
\bibliographystyle{apsrev4-1}

\end{document}